\documentclass[preprint,12pt,3p]{elsarticle}

\usepackage{graphicx}

\usepackage{amssymb}

\usepackage{lineno}

\biboptions{comma,sort&compress}

\usepackage{threeparttable}

\usepackage{lscape}

\renewcommand{\d}[1]{\ensuremath{\mathrm{d}{#1}}}

\journal{Journal of Electroanalytical Chemistry}

\begin{document}

\begin{frontmatter}

\title{Double layer in ionic liquids: capacitance vs temperature}

\author[label1]{Vladislav B. Ivani\v{s}t\v{s}ev}
\address[label1]{Institute of Chemistry, University of Tartu, Ravila 14a, Tartu 50411, Estonia}
\ead{Vladislav.Ivanistsev@ut.ee}

\author[label2]{Kathleen Kirchner}
\address[label2]{Department of Physics, Scottish Universities Physics Alliance (SUPA), Strathclyde University, John Anderson Building, 107 Rottenrow East, Glasgow G4 0NG, U.K.}

\author[label2,label3]{Maxim~V.~Fedorov}
\address[label3]{Skolkovo Institute of Science and Technology, Skolkovo Innovation Center, Moscow 143026, Russia}
\ead{M.Fedorov@skoltech.ru}

\begin{abstract}
Temperature dependence of the capacitance of the electrical double layer (EDL) in molten salts and ionic liquids has been under debates for decades. To rationalise the dependence, we run molecular dynamics simulations of the EDL at variable temperatures. We show that the increase of temperature causes a smearing of the EDL multilayer structure. To explain the results, we propose an~ionic bilayer model. Furthermore, we compare the results with the recent experimental data.
\end{abstract}

\begin{keyword}
ionic liquids \sep temperature dependence \sep electrical double layer \sep capacitance \sep interfacial structure \sep molecular dynamics simulations
\end{keyword}

\end{frontmatter}


\section{Introduction}

Since the 60--70s there have been debates on the nature of the potential and temperature dependence of the electrical double layer (EDL) differential capacitance ($C$) in molten salts \cite{Graves1970,Schmickler1986,Henderson1992,Boda1999a}. 
The recent wave of interest towards the electrochemical applications of ionic liquids (ILs) provoked a~series of new experimental and computational studies \cite{Alam2007wbp,Silva2008,Lockett2008,Lockett2010a,Costa2010,Alam2011,Gnahm2011,Roling2011upb,Druschler2012,Cannes2013ujh, Ivanistsev2013vwi,Siinor2013vgq, Costa2013tqm, Gomes2014txa, Costa2014vks,Loth2010,Vatamanu2010,Dou2011,Kislenko2013wgo,Vatamanu2014uro,Liu2014tvb,Costa2015vkn,Chen2017wlo}. 
Despite remarkable progress achieved in the understanding of the capacitance vs potential dependence \cite{Fedorov2014uyy,Kornyshev2014vjv,Merlet2014vdh}, there is no general agreement on the capacitance vs temperature dependence.

The first studies of the $C(T)$-dependence in ILs showed the increase of capacitance with increasing temperature (positive $\d{C}/\d{T}$ gradient)  \cite{Silva2008,Lockett2008,Lockett2010a,Costa2010,Alam2011}. Later Dr{\"u}schler \textit{et al.} demonstrated that the capacitance can decrease with increasing temperature (negative $\d{C}/\d{T}$ gradient) \cite{Roling2011upb,Druschler2012}. Moreover, Dr{\"u}schler \textit{et al.} suggested that the previously reported strong temperature dependence might be affected by choice of the data analysis. Most recent works demonstrate positive $\d{C}/\d{T}$ gradient in the vicinity of the potential of zero charge (PZC) \cite{Cannes2013ujh,Ivanistsev2013vwi,Siinor2013vgq}, and reveal negative $\d{C}/\d{T}$ gradient at the capacitance peak potential \cite{Costa2014vks,Costa2015vkn}.

\begin{landscape}
\begin{table*} 
\caption{\label{tab:literature_overview}Publications on the temperature dependence of the EDL capacitance. $\d{C_\mathrm{PZC}}/\d{T}$ is the gradients of the dependence at the open circuit or zero charge potential (PZC), and  $\d{C_\mathrm{peak}}/\d{T}$ -- at the potential of the capacitance peak. Positive and negative gradients are marked as ``$+$'' and ``$-$''. ``$\pm$'' sign hints that the dependence is slightly positive or negligible, \textit{i.e.} in the order of the measurement error. ``$\div$'' sign indicates that there are no distinct peaks at the $C(U)$ curve, yet in the whole measured potential range the capacitance increases with increasing the temperature. EIS is an abbreviation of the electrochemical impedance spectroscopy, MC -- Monte Carlo, and MD -- molecular dynamics. In the abbreviations of the ILs, B is for butyl, D -- dodecyl, E -- ethyl, H -- hexyl, Im -- imidazolium, M -- methyl, O -- octyl, P -- pentyl, Pyr -- pyrrolidinium. FEP = tris(pentafluoroethyl)trifluorophosphate, FSI = bis(fluorosulfonyl)imide, TFSI = for bis(trifluoromethane)sulfonimide.}
\begin{threeparttable}
\begin{tabular}{c|l|c|l|l|c|l|l} \hline
Year & References & Method & Ionic Liquids & Electrodes & $T$\,/K & $\displaystyle\frac{\d{C_\mathrm{PZC}}}{\d{T}}$ & $\displaystyle\frac{\d{C_\mathrm{peak}}}{\d{T}}$
\\ \hline \hline
2007 & Alam \textit{et al.} \cite{Alam2007wbp} & EIS & [EMIm][BF$_4$] & Hg & 295--353 & $\pm$ & $-$  \\
2008 & Silva \textit{et al.} \cite{Silva2008} & EIS & [BMIm][PF$_6$] & GC,Hg,Pt & 293--348 & $+$ & $\div$ 
\\ 2008 & Lockett \textit{et al.} \cite{Lockett2008} & EIS & [(E,B,H)MIm][Cl] & GC & 353--413 & $+$ & $+$ \\ 
2010 & Lockett \textit{et al.} \cite{Lockett2010a} & EIS & [(E,B,H)MIm][Cl,Br,I,BF$_4$,PF$_6$,TFSI] & Au,GC,Hg,Pt & 296--373 & $+$ & $+$ \\ 
2010 & Costa \textit{et al.} \cite{Costa2010} & EIS & [(E/B/H/B)MPyr]TFSI & Hg & 293--333 & $+$ & $+$ \\
2011 & Alam \textit{et al.} \cite{Alam2011} & EIS & [(E,B,O)MIm][BF$_4$] & Au(111) & 296--348 & $+$\tnote{a}  & $\div$ \\ 
2011 & Gnahm \textit{et al.} \cite{Gnahm2011} & EIS & [BMIm][PF$_6$] & Au(100) & 293--393 & $\pm$\tnote{b} &\\ 
2012 & Dr{\"u}schler \textit{et al.} \cite{Roling2011upb,Druschler2012} & EIS & [BMPyr][FEP] & Au(111) & 273--363 & $\pm$\tnote{c} & $-$ \\
2013 & Cannes \textit{et al.} \cite{Cannes2013ujh}& EIS & [BMIm][TFSI] & GC,Pt & 298--329 & $+$ & $\div$ \\
2013 & Ivani\v{s}t\v{s}ev \textit{et al.} \cite{Ivanistsev2013vwi}& EIS & [EMIm][BF$_4$] & Cd(0001) &  303--343 & $+$ & $\div$ \\
2013 & Siinor \textit{et al.} \cite{Siinor2013vgq}& EIS & [(E,B)MIm][BF$_4$] & Bi(111) & 298--338 & $+$ & $\div$ \\
2013 & Costa \textit{et al.} \cite{Costa2013tqm}& EIS & [(E,H,D)MIm][TFSI] & Hg &  293--333 & $\pm$ & $\pm$ \\
2014 & Gomes \textit{et al.} \cite{Gomes2014txa}& EIS & [(E,B,H)MIm][BF$_4$,PF$_6$,TFSI] & Au,Pt &  303--333 & $+$ & $+$ \\
2014 & Costa \textit{et al.} \cite{Costa2014vks}& EIS & [(PMIm)$_2$][TFSI]$_2$ & Hg &  313--353 & $\pm$ & $-$ \\
2015 & Costa \textit{et al.} \cite{Costa2015vkn}& EIS & [EMIm][TFSI,FEP] & Hg &  293--333 & $\pm$ & $-$ \\

\hline

2010 & Loth \textit{et al.} \cite{Loth2010} & MC & restricted primitive model & Metal surf. & red. u. & $-$ & \\ 
2010 & Vatamanu \textit{et al.} \cite{Vatamanu2010} & MD & [BMPyr][TFSI] & C(0001) & 363--453 & $\pm$ & $-$ \\ 
2014 & Vatamanu \textit{et al.} \cite{Vatamanu2014uro} & MD & [BMPyr][FSI] & C(0001) & 363--533 & $\pm$ & $-$ \\
2014 & Liu \textit{et al.} \cite{Liu2014tvb} & MD & [BMIm][PF$_6$] & C(0001) & 450--600 & $\pm$ & $-$ \\
2017 & Chen \textit{et al.} \cite{Chen2017wlo} & MD & [EMIm][TFSI] & C(0001) & 350--600 & $\pm$ & $-$ \\
\hline

\end{tabular}

\begin{tablenotes}
\item[a]{$\d{C}/\d{T}$ at the PZC is negative for [EMIm][BF$_4$] and positive for [BMIm,OMIm][BF$_4$].}
\item[b]{Although the capacitance vs temperature was shown to be slightly positive, Gnahm \textit{et al.} concluded that the dependence is negligible \cite{Gnahm2011}.}
\item[c]{Vu \textit{et al.} measured the potential of zero total charge (PZTC) for the same system \cite{Vu2012wkc}. By comparing Refs.~\citenum{Vu2012wkc} and \citenum{Druschler2012}, we take that the negative $\d{C}/\d{T}$ corresponds to the capacitance peak position. Herewith, the dependence at the PZTC is complicated by the surface reconstruction process.}
\end{tablenotes}
\end{threeparttable}
\end{table*}
\end{landscape}

Table~\ref{tab:literature_overview} summarises previous experimental and computational findings. 
We note that most of the experimental measurements were conducted in a~narrower potential range than the simulations, while the computational studies were performed at higher temperatures than the experiments. 
Overall, there are several contradicting explanations of the $C(T)$-dependence which were vividly discussed in Refs.~\citenum{Lockett2010a,Druschler2012,Ivanistsev2013vwi,Vatamanu2014uro,Chen2017wlo}. 

To get more insights into the problem, we have performed molecular dynamics (MD) simulations of a~generic coarse-grained IL confined between two oppositely charged surfaces. The coarse-grain approach was deliberately chosen to avoid the excessive complexity related to molecular structure of the ions. Using the constant surface charge ($\sigma$) method, we evaluated the potential drop ($U$) and the differential capacitance for a~range of temperatures from 250 to 500\,K. This paper presents these results and explains the obtained capacitance vs temperature dependency.

\section{Methods}

\subsection{Simulations details}

MD simulations were run with Gromacs 4.5.5 \cite{Hess2008}. Initial system preparation was performed with the Packmol software \cite{Martinez2009}.  Five independent molecular configurations were generated per each system setup. 34 absolute surface charge ($\sigma$) values were applied. Six different temperatures of 250, 300, 350, 400, 450, and 500\,K were studied. For each system setup, several energy minimisation steps and short structure equilibration steps with a~small time step of $dt = 0.002\,\mathrm{ps}$ were performed. The resulting configurations were used to perform production simulations with $dt = 0.01\,\mathrm{ps}$ in the $NVT$ ensemble. The total number of runs summed up to $34 \times 5 \times 6=1020$, with a~total simulation time for the production runs of $25.5\,\mathrm{\mu s}$. Long simulations resulted in smooth capacitance curves, allowing us to resolve the secondary peaks at the left wing of the capacitance curves in Fig.~\ref{fig:FIG1} ($-2.5\,\mathrm{V} < U < -0.5\,\mathrm{V}$).

Periodic boundary conditions were applied in all directions. The cutoff of the Lennard-Jones interactions was taken to be 2.6\,nm with the shifted potential method to account for the coarse-grain model of ions. The long-range Coulomb interactions were handled by the particle-mesh Ewald method with a~cutoff of 2.9\,nm and a~grid spacing of 0.112\,nm and corrected for slab geometry \cite{Essmann1995,Yeh1999}. The neighbour list for non-bonded interactions was updated every 10th integration step. The time step of 0.01\,ps was used in the leap-frog algorithm for integrating Newton's equations of motion. All simulations were performed at fixed temperatures. Velocity re-scaling was used with a~temperature coupling constant of 1.0\,ps \cite{Bussi2007}.

All~simulations were automated using the scripting framework NaRIBaS (Nanomaterials and Room-temperature Ionic liquids in Bulk and Slab) \cite{Ivanistsev2015tvg}.

\subsection{Model details}
A square lattice of 2500 Lennard-Jones spheres atoms was chosen to model the surfaces with $11\,\mathrm{nm}\times 11\,\mathrm{nm}$ size in $x$ and $y$ directions. The position of the surface atoms was restrained. The distance between the two surfaces was set to $24\,\mathrm{nm}$ at 450~K and was varied with the temperature. The ion pair number was fixed to 1050 in all simulations. a~vacuum slab prolonged the box size in $z$-direction to 40\,nm, thereby following the Berkowitz rule to obtain correct electrostatics simulations in slab geometry \cite{Yeh1999}. 
The surfaces were charged oppositely with $\sigma$ ranging from 0 to $\pm 50\,\mathrm{\mu C/cm^2}$. A~surface charge of $1\,\mathrm{\mu C/cm^2}$ results in a~``physical'' charge per atom of $3.021 \cdot 10^{-3}$ elementary charges. The effective charge used in the simulations was obtained by scaling the ``physical'' charge by $1/\sqrt2$. As we use a primitive IL model, the denominator corresponds to the high-frequency permittivity of 2 that is closer to the optical permittivity of halides rather than more complex ILs. 

The coarse-grain model of the IL was taken from Ref.~\citenum{Fedorov2008a}. Ions represent charged Lennard-Jones spheres with a short-range repulsive pairwise potential preventing the crystallisation.  The model is not supposed to mimic any real IL, although, concerning ionic size, it is similar to trihexyl(tetradecyl)-phosphonium tris(pentafluoroethyl) trifluorophosphate. The modelling of Lennard-Jones spheres allows to “separate wheat from chaff” by eliminating the effect of complex shapes packing and focusing on the primary impact of ionic layer formation.

\section{Results and discussions}

\subsection{Calculated differential capacitance $C(U,T)$}

\begin{figure*}[ht]
\begin{center}
\includegraphics[scale=1]{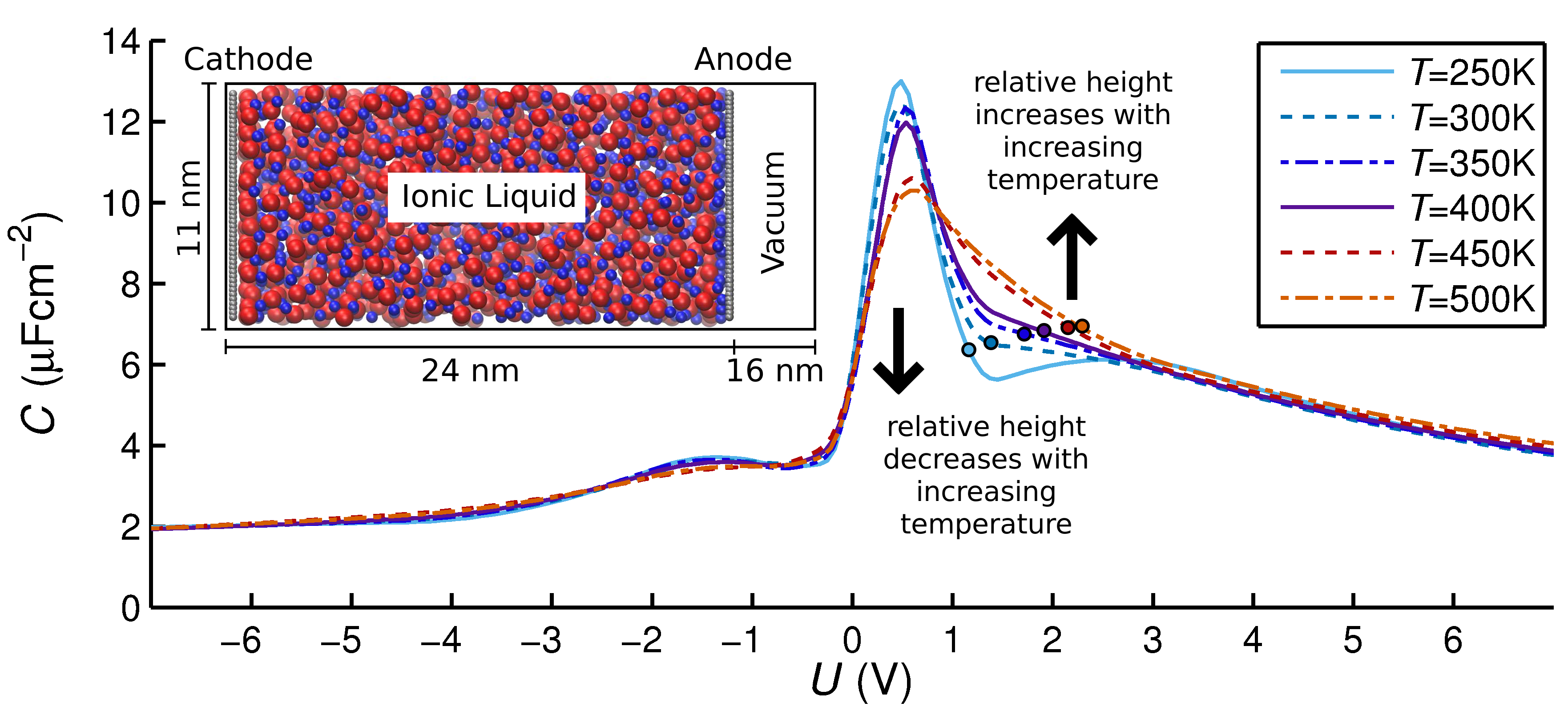}
\end{center}
\caption{\label{fig:FIG1} Differential capacitance ($C$) as a~function of the potential drop ($U$) for six different temperatures ranging from 250\,K to 500\,K. The PZC equals $-0.05$\,V and is almost temperature-independent. The capacitance values corresponding to the maximal cumulative number differences in~Fig.~\ref{fig:FIG3}(d) are shown with circles. The error bars for the data noise of the capacitance calculations (not shown) steadily increase with the voltage from $\pm0.1\,\mathrm{\mu F/cm^2}$ to $\pm1.0\,\mathrm{\mu F/cm^2}$ (see Appendix for details). The inset on the left sketches the geometry of the simulation cell in $xy$-plane. Cations are shown as red spheres, anions are shown as blue spheres, and surface atoms are shown as grey spheres. The size of the particles is rescaled to make them more visible. The sketch is illustrative and does not reproduce the proper scale.}
\end{figure*}

Fig.~\ref{fig:FIG1} shows the $C(U)$ curves at six different temperatures. 
At very low and very high potential values, where $C(U,T)$ wings merge, the capacitance is temperature independent. 
Within $-2.5\,\mathrm{V} < U < 2.5\,\mathrm{V}$ range, the sign of the $\d{C}/\d{T}$ gradient changes from negative at the $C(U)$ peak potentials to positive at the PZC. 
Similarly,  in experiments either negligible or positive $\d{C}/\d{T}$ gradient is observed in the vicinity of the PZC \cite{Lockett2008,Lockett2010a,Gnahm2011,Druschler2012,Alam2011,Silva2008,Costa2010,Cannes2013ujh,Ivanistsev2013vwi,Siinor2013vgq}. Recent experiments also reveal a~clearly negative $\d{C}/\d{T}$ gradient at the $C(U)$ peak potential \cite{Costa2014vks, Costa2015vkn}.

The most distinct temperature effect is seen at +0.5\,V, around the $C(U)$ peak. The height of the peak \emph{decreases} with increasing the temperature from $13\,\mathrm{\mu F/cm^2}$ at 250\,K to $10\,\mathrm{\mu F/cm^2}$ at 500\,K. This decrease is accompanied by \emph{widening} of the peak and slight steady \emph{shift} of its position towards positive potentials. The $\d{C}/\d{T}$ gradient changes from negative at the peak potential to positive at
potentials higher than +0.8\,V.

\subsection{Theoretical models}

In the literature one can find several explanations of the $C(T)$ dependence. The capacitance was repeatedly expressed to explain both experimental and computational results in Refs.  \cite{Alam2007wbp,Silva2008,Lockett2008,Lockett2010a,Costa2010,Siinor2013vgq,Costa2013tqm, Gomes2014txa, Costa2014vks, Costa2015vkn,Vatamanu2014uro,Liu2014tvb} as follows:

\begin{equation}
\label{eq:Helmholtz}
C \simeq \frac{\varepsilon_\mathrm{i} \varepsilon_0}{l},
\end{equation}

\noindent where $l$ is the distance from the surface layer to the mass-centre of counter-ion layer, $\varepsilon_\mathrm{i}$ is an interfacial dielectric constant, and $\varepsilon_0$ is the vacuum permittivity. On the one hand, $\varepsilon_\mathrm{i}$ in Eq.~\ref{eq:Helmholtz} should increase up to the bulk values ($\varepsilon_\mathrm{b}$) with increasing temperature, as heating reveals more degrees of freedom. On the other hand, according to the Kirkwood's formula, $\varepsilon_\mathrm{b}$ decreases with increasing temperature. For these reasons, the $\d{C}/\d{T}$ gradient could be positive while $\varepsilon_\mathrm{i} \rightarrow \varepsilon_\mathrm{b}$, then it should change to negative.

In the mean spherical approximation theory accounting for mass-action law (MSA-MAL) \cite{Holovko2001whu,Boda1999a}, the capacitance is approximated as:

\begin{equation}
\label{eq:Boda}
C \simeq \frac{\varepsilon_\mathrm{i} \varepsilon_0}{\lambda_\mathrm{D}}\sqrt{\alpha},
\end{equation}

\noindent where $\lambda_\mathrm{D}$ is the Debye length, and $\alpha$ is a~dissociation constant. The pure MSA theory predicts $C \sim T^{-1/2}$. When an association of ions in IL is taken into account via the mass-action law, the capacitance increase can take place as $\alpha$ increases with increasing temperature. The MSA-MAL theory predicts that the $\d{C}/\d{T}$ gradient may be positive while $\alpha \rightarrow 1$, then it changes to negative. A similar narrative was recently used by Chen \textit{et al.} to extend the lattice-gas mean-field theory of ILs \cite{Chen2017wlo}.

A different approach is used in the model of ``counter-charge layer in generalised solvents'' \cite{Feng2011vvj}, where the (integral) capacitance is expressed as:

\begin{equation}
\label{eq:Feng}
C \simeq \frac{\varepsilon_\mathrm{\infty} \varepsilon_0}{l+ \sum_{i=1}^N(-1)^i\gamma_i\Delta_i},
\end{equation}

\noindent where $\varepsilon_\infty$ is the high-frequency dielectric constant, $\gamma_i$ is a~screening constant, and $\Delta_i$ is an average distance between the counter- and co-ion of the $i$-th layer (see Ref. \citenum{Feng2011vvj} for details).  Eq.~\ref{eq:Feng} simplifies to Eq.~\ref{eq:Helmholtz} by defining $\varepsilon_\mathrm{i} = \varepsilon_\infty/[{1 + \sum_{i=1}^N(-1)^j\gamma_i\Delta_i/l}]$. The so-called ``melting'' of the layered structure due to the increase of temperature may hypothetically lead to a~decrease of the number of layers ($N$). Numerically, for an even $N$ the capacitance is higher than for an odd $N$. Thus, under thermal distortion, the disappearance of layers may result in both negative and positive $\d{C}/\d{T}$ gradient.

The interpretations accounting for relative dielectric constant ($\varepsilon_\mathrm{i}$), dissociation constant ($\alpha$), and screening constant ($\gamma$) provide a~similar qualitative description. For instance, it is possible to speculate that the ion association is thermally disrupted and the EDL thins or widens. Qualitatively it allows to refer Eq.~\ref{eq:Helmholtz} while discussing the temperature dependence. Such interpretations treat the EDL in ILs essentially as a~combination of a~diffuse layer and a~Helmholtz layer, in-spite experimentally confirmed multilayer structure. 
Eq.~\ref{eq:Boda} was previously used in discussions of the ``anomalous'' positive $\d{C}/\d{T}$ gradient at the PZC, where is should be negative according to the Gouy--Chapman theory \cite{Holovko2001whu,Boda1999a}. 
Our simulations reveal a~very small positive $\d{C}/\d{T}$ gradient at the PZC which is within the error estimates. 
The positive $\d{C}/\d{T}$ gradient at potentials higher than +0.8\,V is not related to the ``anomalous'' $C(T)$ dependence, as it is far from the PZC (see Fig.~\ref{fig:FIG1}).
For this reasons, in this paper, we avoid speculating on the ``anomalous'' temperature dependence. 
Instead, we focus on the negative $\d{C}/\d{T}$ gradient at the peak potential that is related to the EDL restructuring upon charging.

\subsection{Structural variations upon heating}

\begin{figure*}
\centering
\includegraphics[width=\textwidth]{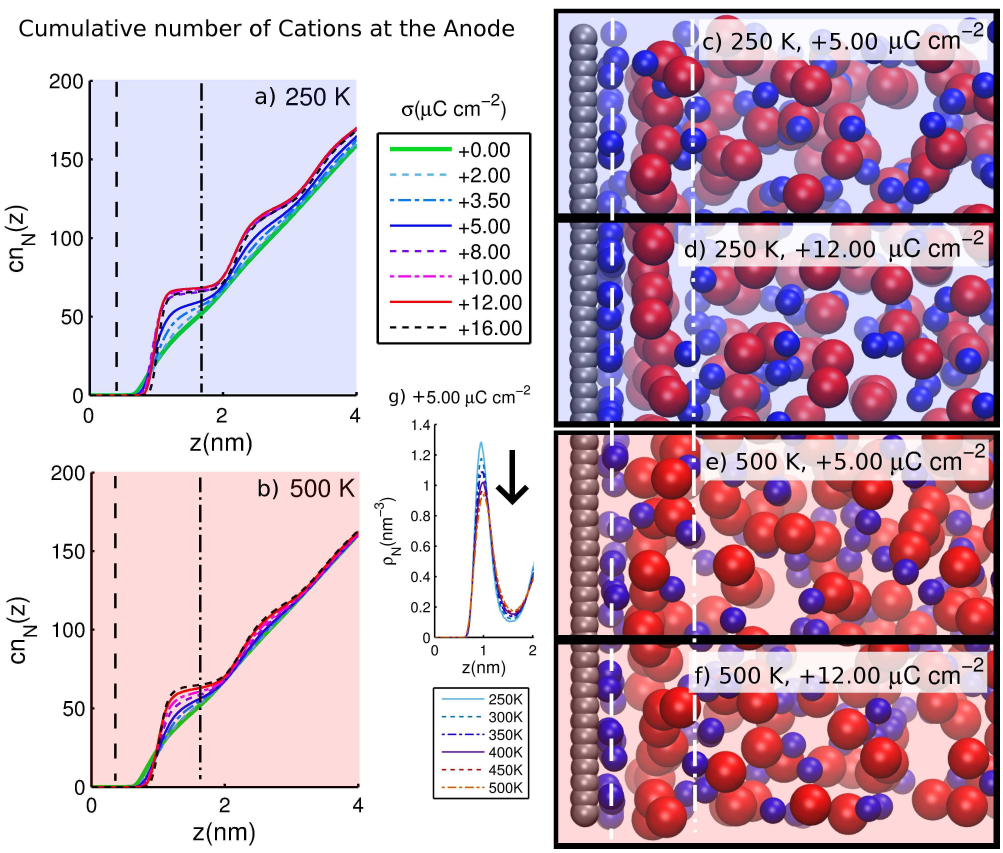}
\caption{\label{fig:FIG2} (LEFT) Simulated cumulative number curves of cations at the positively charged surface for different surface charges between $0\,\mathrm{\mu C/cm^2}$ ($U = -0.1$\,V) and $+16\,\mathrm{\mu C/cm^2}$ (+1.8\,V) at two temperatures: (a) 250\,K and (b) 500\,K. The dashed and dot-dashed lines indicate the positions of anion layers. (RIGHT) Simulation snapshots of the interfacial ion configurations at (c) 250\,K and $+5\,\mathrm{\mu C/cm^2} (+0.6\,V)$, (d) 250\,K and $+12\,\mathrm{\mu C/cm^2}$ (+1.2\,V), (e) 500\,K and $+5\,\mathrm{\mu C/cm^2}$ (+0.6\,V), and (f) 500\,K and $+12\,\mathrm{\mu C/cm^2}$ (+1.2\,V). White dashed and dot-dashed lines indicate again the anion layer positions. (g) Number density profile of cations at the positively charged surface for $+5\,\mathrm{\mu C/cm^2}$ comparing different temperatures. Within the volume between the anion layers the cations show a~broad distribution for the low surface charges. At intermediate surface charges and low temperature, a~stiff cation layer is formed with a~well-defined distance from the surface. Upon increasing the temperature, the cation layer is smeared as shown by the broadening of the cation distribution (g) and the smoothed slope of increase of the cumulative number curves (b).}
\end{figure*}

Fig.~\ref{fig:FIG2} highlights the structural differences at variable temperatures in terms of the total number of ions within a~distance $z$ from the surface plane ($\mathrm{cn}_\mathrm{N}(z)$). Note that the steps in the $\mathrm{cn}_\mathrm{N}(z)$ profiles indicate the formation of an ion layer. Smoothing of the $\mathrm{cn}_\mathrm{N}(z)$ profiles (Fig.~\ref{fig:FIG2}(a)--(b)) implies a~smearing process. 

It follows that the smearing of the second, co-ion interfacial layer is of a~greater importance than of the very first, counter-ion layer. Snapshots in Fig.~\ref{fig:FIG2}(c)--(d) depict how this layer is smeared with an increase of temperature. At~positive surface charges, the second layer is formed by cations on-top of the first layer of anions. While at low temperature the cation layer is stiff, it is smeared upon heating. The smearing is reflected also in the broadening of the cation distribution (Fig.~\ref{fig:FIG2}(g)). 

\begin{figure*}
\centering
\includegraphics[width=\textwidth]{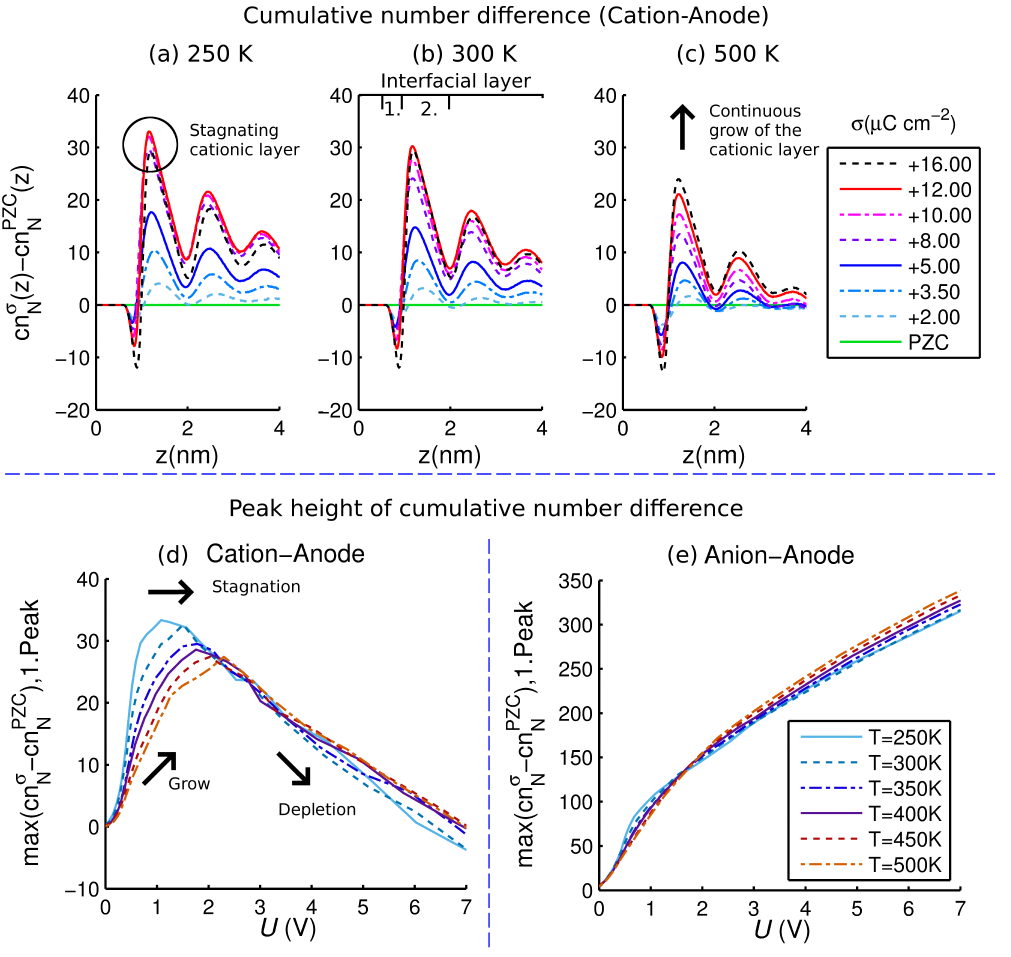}
\caption{\label{fig:FIG3} (TOP) Simulated cumulative number differences of cations -- with PZC curve being the reference curve -- at the positively charged surface for different voltages between +0.2\,V and +1.8\,V. From left to right three different temperatures are considered: (a) 250\,K, (b) 300\,K and (c) 500\,K. The total number of cations increases while the surface gets positively charged. The first minimum of the cumulative number difference can be explained by an increase of the electrostatic repulsion between surface and ions. The comparison of the peak height at the same potential but different temperatures reveals the smearing effect -- the number of cations in the EDL decreases with increasing temperature. However, at low temperatures, we find a~stagnation in the layer formation process, whereas at high temperatures the EDL grows continuously with increasing voltage. (BOTTOM) The height of the first peak of the cumulative number difference over potential drop for (d) cations at the positively charged surface and (e) anions at the positively charged surface for all temperatures between 250\,K and 500\,K. The peak height corresponds to the varying amount of ions in the EDL as a~response to the increased surface charge with respect to the EDL at the neutral surface. For a~discussion see main text.}
\end{figure*}

To rationalise the negative $\d{C}/\d{T}$ gradient at the peak potential, let us focus on cumulative number difference $\mathrm{cn}^\sigma_\mathrm{N}(z)-\mathrm{cn}^\mathrm{PZC}_\mathrm{N}(z)$. Fig.~\ref{fig:FIG3}(a)--(c) illustrates the difference at temperatures of 250\,K, 300\,K, and 500\,K. 
The minimum in the cumulative number difference is due to the repulsion of cations by the positive surface charge -- with increasing the surface charge, the number of cations in the first layer decreases until all cations leave.
The maximum in the cumulative number difference is due to the accumulation of cations in the second layer.
As can be seen in Fig.~\ref{fig:FIG3}(a)--(c), with increase of temperature, the height of these maxima decrease.

Fig.~\ref{fig:FIG3}(d)--(e) shows the potential and temperature dependencies of the height of the first maximum for counter-ion (anion) and co-ion (cation) layers. 
As can be seen in Figs.~\ref{fig:FIG3}(d), the second layer reaches a~state of maximal number density ($\theta_2$). 
That state corresponds to the dense layer of \emph{co-ions}. 
Previously, we introduced the \emph{counter-ions} monolayer ansatz \cite{Kirchner2013tss,Ivanistsev2014wtw,Ivanistsev2015usp}. 
For the given model, the potential of the counter-ion monolayer formation is reached at very high values (not shown in Fig.~\ref{fig:FIG3}(e)); it is insensitive to the temperature and corresponds to a state with $\theta_1=\theta_\mathrm{max}$ and $\theta_2=0$. 

The concept of dense co-ion layer with $\theta_2$ is similar to the monolayer concept, as both are determined by the packing of ions. 
However, the maximal density of the co-ion is strongly dependent on the temperature -- in Fig.~\ref{fig:FIG3}(d), for $U < 1.8\,\mathrm{V}$, the number of cations decreases with increasing temperature.

Due to the thermal distortion, the formation of the dense co-ion layer is delayed on the potential scale. At 250\,K it forms around 1.2\,V, while at 500\,K -- around 2.4\,V. The stagnation of the co-ion layer formation as well as the delay of the stagnation are the most significant observations of this work. To our best knowledge, they have not been reported, although the former was briefly discussed in Ref.~\citenum{Ivanistsev2015usp}. To explain the phenomena we propose a~model of the EDL represented as an ionic bilayer.

\subsection{Ionic bilayer model}

The ionic bilayer (IBL) model divides the EDL into three regions: surface, IL bulk and an ionic bilayer -- a~layer of counter-ions plus a~layer of co-ions \cite{Ivanistsev2014tqg}. Accordingly, with increase of the surface charge 1) counter-ions from the bulk can accumulate in the first layer, 2) co-ions can be repelled from the first to the second layer, 3) co-ions can be repelled from the second layer to the bulk. 
The filling of the second layer with co-ions is part of the EDL charging mechanism \cite{Ivanistsev2014tqg,Ivanistsev2015usp}. 
In general, the EDL charging can happen through 1) counter-ion attraction and 2) co-ion repulsion from the EDL. 
Considering the first layer, these processes correspond to adsorption and desorption, respectively. 
The third possible mechanism imply co-ion/counter-ion exchange. 

Due to electroneutrality $q_1\mathrm{cn}_\mathrm{1} + q_2\mathrm{cn}_\mathrm{2} = -\sigma A$, where $A$ is an area, ${q}_i$ is ionic charge, and $\mathrm{cn}_i$ is the number of ions in the $i$-th layer.
In the IBL model, the ionic charge density of the second layer is the excess charge density, which is equal and opposite to the overscreening charge density of the first layer. 
Let us denote $\mathrm{cn}_\mathrm{2}/A$ as $\lambda$.

As the EDL represents a~parallel plate capacitor, the potential drop across the IBL model can be expressed as:

\begin{equation}
\label{eq:Udrop}
U = -{\frac{\sigma}{\varepsilon_\infty\varepsilon_0} \frac{\int_0^\infty z\rho_\mathrm{ion}(z)\d{z}}{\int_0^\infty \rho_\mathrm{ion}(z)\d{z}}} = \frac{l\sigma - \delta\lambda}{\varepsilon_\infty\varepsilon_0},
\end{equation}

\noindent where the surface charge plane is situated at $z=0$, while the mass-centre of the ionic charge density is defined by the positions of the first ($l$) and the second ($l + \delta$) layers. Notably, from Eq.~\ref{eq:Udrop} follows that the highest possible overscreening value ($\lambda/\sigma+1$) is simply defined by the ratio $l/\delta$ as $\lambda/\sigma+1 < l/\delta+1$.

In our simulations there is no indication of significant widening or narrowing of the EDL upon heating. Dou \textit{et al.} and Nishi \textit{et al.} made the same observation using MD simulations and X-ray reflectivity, respectively \cite{Dou2011,Nishi2011vpv}. For this reason, the layer positions are taken to be rigid; \textit{i.e.} $l$ and $\delta$ are potential- and temperature-independent. Because $l\sigma = \varepsilon_\infty\varepsilon_0 U + \delta\lambda$, the differential capacitance of our simplistic model is expressed as:

\begin{equation}
\label{eq:IBLC}
C = \frac{\d \sigma}{\d U} = \frac{\varepsilon_\infty\varepsilon_0}{l} + \frac{\delta}{l} \nabla\lambda,
\end{equation}

\noindent where $\nabla\lambda = {\d \lambda}/{ \d U}$. When Eq.~\ref{eq:IBLC} simplifies to Eq.~\ref{eq:Helmholtz}. In comparison to the multilayer EDL seen in the simulations, the assumption that $l$ is potential independent breaks above the potential of the monolayer formation. The assumption that $\delta$ is potential independent is invalid near the PZC where counter- and co-ionic layers are not yet segregated. 
Still, using the IBL model it is possible to give phenomenological explanation of both potential and temperature dependencies presented in Fig.~\ref{fig:FIG1}.

\subsection{Phenomenological explanation}

Let us assume that $\lambda \sim$ in Eq.~\ref{eq:IBLC} is proportional to $\max{(\mathrm{cn}^\sigma_\mathrm{N}(z)-\mathrm{cn}^\mathrm{PZC}_\mathrm{N}(z))}$ from Fig.~\ref{fig:FIG3}(d). 
Then the changes in the gradient $\nabla\lambda$ can be related to the existence of a~$C(U)$ peak (Fig.~\ref{fig:FIG1}) at maximal $\nabla\lambda$ as well as the stagnation of the second layer (Fig.~\ref{fig:FIG2}) at $\nabla\lambda = 0$.

The position of the $C(U)$ peak is defined by the maximal rate of co-ion accumulation in the second layer (maximal $\nabla\lambda$). 
This rate can be deduced from the slope in Fig.~\ref{fig:FIG3}(d). 
As the slope reduces with increasing the temperature, the peak capacitance also decreases.
The reduction of the slope means that the smearing of the EDL suppresses both overscreening and excess charge stored in the multilayer structure.

The maximal charge excess in the second layer ($\theta_2$) is determined by the packing on ions.
The position of the maxima in Fig.~\ref{fig:FIG3}(d) shifts towards higher potentials and its height decreases with increasing the temperature. 
Also, from Fig.~\ref{fig:FIG2}(a)--(b) it follows that, in accordance with Eq.~\ref{eq:Udrop}, at higher temperatures $\theta_2$ is reached at higher surface charges. 
In~other words, due to the smearing of the whole EDL the stagnation is delayed. 
At this specific point $\nabla\lambda = 0$ and the capacitance is simply $C = {\varepsilon_\infty \varepsilon_0}/{l}$. 
As can be seen in Fig.~\ref{fig:FIG1}, it is almost temperature independent. 

The presented results are in remarkable agreement with the recent experimental data from Refs.~\citenum{Costa2014vks,Costa2015vkn}, where the negative $\d{C}/\d{T}$ gradient was observed at the capacitance peak potential. 
To verify, whether the proposed explanations are valid, we call upon more detailed experimental studies of capacitance vs temperature dependence. 
We would like to turn attention to the interfaces showing the so-called ``camel''-shape $C(U)$ dependency \cite{Wallauer2013uvg,Costa2015tae,Oll2017wes}. 
For such interfaces, it is possible to determine the $\d{C}/\d{T}$ gradient at the PZC as well as at the capacitance peak potential.
Modern spectroscopy and microscopy methods could be used to confirm or refute the stagnation phenomena \cite{Romann2014uqi,Atkin2014tjb,Mezger2015tdg,Oll2016tjn}. 
Also, the potential of maximum entropy could be determined from laser-induced heating \cite{Sebastian2015uzi}.

\section{Conclusions}

To sum up, the following conclusions apply for the temperature effects on the EDL in the coarse-grained IL:

\begin{itemize}
\item[1)] Overall, the temperature effect on the EDL capacitance ($C$) depends on the potential. 
\item[2)] At large absolute potentials (higher than $2.5$ V), the structure of the EDL is determined by the strong electrostatic \emph{surface--counter-ion} interactions and the packing of counter-ions. It results in very densely packed interfacial structures (see also \cite{Ivanistsev2014tqg,Ivanistsev2015usp}) and very weak dependence of the capacitance on temperature.

\item[3)] At lower absolute potentials, the structure of the EDL is determined by the interplay between the electrostatic \emph{anion--cation} interactions and the packing of ions. The first one is responsible for the overscreening, while the second one causes the stagnation of the EDL layering which limits the overscreening. 

\item[3.1)] The increase of temperature suppresses the overscreening. It leads to the decrease of capacitance at its peak potential (negative $\d{C}/\d{T}$ gradient).

\item[3.2)] Heating also causes smearing of the EDL layering. It leads to the delay of stagnation of the EDL layering on the potential scale, which results in a~positive $\d{C}/\d{T}$ gradient for moderate potentials.
\end{itemize}

Earlier we proposed that the overscreening can be expressed using the charge excess ($\lambda$). We concluded that the maximal EDL layering coincides with the stagnation of the co-ion layer \cite{Ivanistsev2014tqg,Ivanistsev2015usp}. In this work, we have shown that the stagnation happens when $\nabla\lambda = 0$ as well as that the capacitance peak corresponds to the maximal gradient $\nabla\lambda$. According to the proposed ionic bilayer model, at the stagnation potential, the capacitance is almost temperature-independent. On the opposite, at the $C$ peak potential, the capacitance decreases with increasing the temperature. The simulations results confirm these predictions.

\section{Appendix A. Calculation of the differential capacitance and error estimation}

The analysis of the obtained trajectories was performed using Gromacs tools and self-written Matlab functions.

The differential capacitance was obtained while approaching the following steps for all trajectories: 1) Calculation of the number density profiles $n(z)$ for cations and anions respectively using a~uniform grid with a~spacing of 0.015\,nm; 2) Calculation of the charge density profile by summing the charge scaled number density profiles $\rho_\mathrm{ion}(z)=\sum q  n_\mathrm{cation}(z) - \sum q n_\mathrm{anion}(z)$ with $q=1e$; 3) Integration of the charge density profiles multiplied by $z$ to obtain the potential drop: $U=\int z \cdot \rho_\mathrm{ion}(z)\d z$; 4) The resulting surface charge ($\sigma$) vs potential drop ($U$) curves were first extended by the expected behaviour at high surface charges (therefore we calculated the least square fit of $a\sqrt{U}+b$ to the asymptotic region); afterwards we smoothed the extended data set using piece-wise polynomial curve fitting by weighted least squares; 5) Finally the smoothed curves are differentiated to obtain the differential capacitance $ C= \d{\sigma} / \d{U}$.

The error of the differential capacitance needs to be estimated as the combined error of all preparation steps that finally result in the capacitance curve. The steps that can be considered of major importance in this error analysis are the quality of the potential drop data and the assumption made for asymptotic extrapolation. The first error source is the noise in the data. The asymptotic extrapolation introduces a systematic error. Therefore the numerical differentiation error is negligible. 

We assumed that our data set $f(x)$ is approximated by a~fit function $g(x)$ with a~\emph{data noise} $e(x)$, which we considered being the 95\,\% confidence interval of the surface charge $\sigma$ vs potential drop $U$ curves. To obtain the data noise of the first derivative of $f(x)$, the error of the fit function was differentiated $e''(x)=\frac{\mathrm d e}{\mathrm d x}=e(f''(x))$.

Differentiating the confidence interval results in an error of 0.4\,\% for the maximum equal to $C=0.05\,\mathrm{\mu C/cm^2}$. At the boundaries ($U=\pm7.0\,\mathrm V$) we estimated an error of 4.7\,\% equal to $C=0.09\,\mathrm{\mu C/cm^2}$ at the negative axis and an error of 2.0\,\% equal to $C=0.1\,\mathrm{\mu C/cm^2}$ at the positive axis. The absolute errors are equal for all temperatures.

The estimation of the \emph{systematic error} introduced by the fitting assumption for the asymptotic wings is more complicated. The systematic error is introduced by the asymptotic extrapolation of the surface charge vs potential drop data. The assumption was made that the asymptotes follow a~$\left|U\right|^{0.5}$ dependency as predicted by using fundamental principles as the charge conservation law \cite{Kornyshev2007}. When fitting the fractional exponent only for the right -- wing of the surface charge vs potential drop curve, the exponent shows a~reliable dependency between $\left|U\right|^{0.66}$ for 250\,K and $\left|U\right|^{0.81}$ for 500\,K. Similar values have been reported earlier by Vatamanu \textit{et al.} \cite{Vatamanu2010}. For the left -- wing the exponent cannot be determined as lattice saturation not reached at +7.5\,V. 

When changing the assumption from a~fractional power law with exponent 0.5 to 0.8, the shape of the differential capacitance changes drastically in the high positive voltage interval. Moreover the minimum in the 250\,K capacitance curve at 1.5\,V vanishes. However, the general temperature dependency including the intersection point at +0.8\,V is not affected by this systematic error.

\section{Acknowledgements}
We thank Prof. A.A. Kornyshev, Prof. G.A. Tsirlina, and Dr. V. Lockett for useful discussions.
We acknowledge the supercomputing support from the von Neumann-Institut f{\"u}r Computing, FZ J{\"u}lich (Project ID ESMI11), the ARCHIE-west supercomputing centre (EPSRC grant no. EP/K000586/1) and the High Performance Computing Center of the University of Tartu. This work was supported by Grant FE 1156/2-1 of the Deutsche Forschungsgemeinschaft (DFG), by the Estonian Personal Research Project PUT1107, and by the EU through the European Regional Development Fund (TK141 ``Advanced materials and high-technology devices for energy recuperation systems''). K.K. thanks the Max Planck Institut f{\"u}r Mathematik in den Naturwissenschaften for hospitality during her stay there and access to the local computing facilities.


\bibliographystyle{elsarticle-num}

\bibliography{References}

\begin{thebibliography}{10}
\expandafter\ifx\csname url\endcsname\relax
  \def\url#1{\texttt{#1}}\fi
\expandafter\ifx\csname urlprefix\endcsname\relax\def\urlprefix{URL }\fi
\expandafter\ifx\csname href\endcsname\relax
  \def\href#1#2{#2} \def\path#1{#1}\fi

\bibitem{Graves1970}
A.~D. Graves, D.~Inman, The electrical double layer in molten salts. part 2.
  the double-layer capacitance, Journal of Electroanalytical Chemistry 25~(3)
  (1970) 357--372.

\bibitem{Schmickler1986}
W.~Schmickler, D.~Henderson, New models for the structure of the
  electrochemical interface, Progress in Surface Science 22~(4) (1986)
  323--420.

\bibitem{Henderson1992}
J.~R. Henderson, Z.~A. Sabeur, Liquid-state integral-equations at high-density
  - on the mathematical origin of infinite-range oscillatory solutions, Journal
  of Chemical Physics 97~(9) (1992) 6750--6758.

\bibitem{Boda1999a}
D.~Boda, D.~Henderson, K.~Y. Chan, D.~T. Wasan, Low temperature anomalies in
  the properties of the electrochemical interface, Chemical Physics Letters
  308~(5-6) (1999) 473--478.

\bibitem{Alam2007wbp}
M.~T. Alam, M.~M. Islam, T.~Okajima, T.~Ohsaka,
  \href{http://dx.doi.org/10.1021/jp075808l}{Measurements of differential
  capacitance at {Mercury/Room-Temperature} ionic liquids interfaces}, Journal
  of Physical Chemistry C 111~(49) (2007) 18326--18333.
\newblock \href {http://dx.doi.org/10.1021/jp075808l}
  {\path{doi:10.1021/jp075808l}}.
\newline\urlprefix\url{http://dx.doi.org/10.1021/jp075808l}

\bibitem{Silva2008}
F.~Silva, C.~Gomes, M.~Figueiredo, R.~Costa, A.~Martins, C.~M. Pereira, The
  electrical double layer at the [bmim][pf6] ionic liquid/electrode interface
  -- effect of temperature on the differential capacitance, Journal of
  Electroanalytical Chemistry 622~(2) (2008) 153--160.

\bibitem{Lockett2008}
V.~Lockett, R.~Sedev, J.~Ralston, M.~Horne, T.~Rodopoulos, Differential
  capacitance of the electrical double layer in imidazolium-based ionic
  liquids: Influence of potential, cation size, and temperature, Journal of
  Physical Chemistry C 112~(19) (2008) 7486--7495.

\bibitem{Lockett2010a}
V.~Lockett, M.~Horne, R.~Sedev, T.~Rodopoulos, J.~Ralston, Differential
  capacitance of the double layer at the electrode/ionic liquids interface,
  Physical Chemistry Chemical Physics 12~(39) (2010) 12499--12512.
\newblock \href {http://dx.doi.org/10.1039/c0cp00170h}
  {\path{doi:10.1039/c0cp00170h}}.

\bibitem{Costa2010}
R.~Costa, C.~M. Pereira, F.~Silva, Double layer in room temperature ionic
  liquids: influence of temperature and ionic size on the differential
  capacitance and electrocapillary curves, Physical Chemistry Chemical Physics
  12~(36) (2010) 11125--11132.
\newblock \href {http://dx.doi.org/10.1039/c003920a}
  {\path{doi:10.1039/c003920a}}.

\bibitem{Alam2011}
M.~T. Alam, J.~Masud, M.~M. Islam, T.~Okajima, T.~Ohsaka, Differential
  capacitance at au(111) in 1-alkyl-3-methylimidazolium tetrafluoroborate based
  room-temperature ionic liquids, Journal of Physical Chemistry C 115~(40)
  (2011) 19797--19804.
\newblock \href {http://dx.doi.org/10.1021/jp205800x}
  {\path{doi:10.1021/jp205800x}}.

\bibitem{Gnahm2011}
M.~Gnahm, C.~M\"{u}ller, R.~R\'{e}p\'{a}nszki, T.~Pajkossy, D.~M. Kolb, The
  interface between au(100) and
  1-butyl-3-methyl-imidazolium-hexafluorophosphate, Physical Chemistry Chemical
  Physics 13~(24) (2011) 11627--11633.
\newblock \href {http://dx.doi.org/10.1039/c1cp20562e}
  {\path{doi:10.1039/c1cp20562e}}.

\bibitem{Roling2011upb}
B.~Roling, M.~Drüschler, B.~Huber,
  \href{http://pubs.rsc.org/en/content/articlelanding/2012/fd/c1fd00088h}{Slow
  and fast capacitive process taking place at the ionic liquid/electrode
  interface}, Faraday Discussions 154 (2011) 303--311.
\newblock \href {http://dx.doi.org/10.1039/C1FD00088H}
  {\path{doi:10.1039/C1FD00088H}}.
\newline\urlprefix\url{http://pubs.rsc.org/en/content/articlelanding/2012/fd/c1fd00088h}

\bibitem{Druschler2012}
M.~Dr\"{u}schler, N.~Borisenko, J.~Wallauer, C.~Winter, B.~Huber, F.~Endres,
  B.~Roling, New insights into the interface between a single-crystalline metal
  electrode and an extremely pure ionic liquid: slow interfacial processes and
  the influence of temperature on interfacial dynamics, Physical Chemistry
  Chemical Physics 14~(15) (2012) 5090--5099.
\newblock \href {http://dx.doi.org/10.1039/c2cp40288b}
  {\path{doi:10.1039/c2cp40288b}}.

\bibitem{Cannes2013ujh}
C.~Cannes, H.~Cachet, C.~Debiemme-Chouvy, C.~Deslouis, J.~de~Sanoit,
  C.~Le~Naour, V.~A. Zinovyeva,
  \href{http://pubs.acs.org/doi/10.1021/jp407665q}{Double layer at
  [{BuMeIm][Tf2N]} ionic {Liquid--Pt} or {C} material interfaces}, The Journal
  of Physical Chemistry C 117~(44) (2013) 22915--22925.
\newblock \href {http://dx.doi.org/10.1021/jp407665q}
  {\path{doi:10.1021/jp407665q}}.
\newline\urlprefix\url{http://pubs.acs.org/doi/10.1021/jp407665q}

\bibitem{Ivanistsev2013vwi}
V.~Ivani{\v s}t{\v s}ev, A.~Ruzanov, K.~Lust, E.~Lust, Comparative impedance
  study of cd(0001) electrode in {EMImBF4} and {KI} aqueous solution at
  different temperatures, Journal of The Electrochemical Society 160~(6) (2013)
  H368--H375.
\newblock \href {http://dx.doi.org/10.1149/2.129306jes}
  {\path{doi:10.1149/2.129306jes}}.

\bibitem{Siinor2013vgq}
L.~Siinor, R.~Arendi, K.~Lust, E.~Lust, Influence of temperature on the
  electrochemical characteristics of bi(111){\textbar}ionic liquid interface,
  Journal of Electroanalytical Chemistry 689 (2013) 51--56.
\newblock \href {http://dx.doi.org/10.1016/j.jelechem.2012.11.018}
  {\path{doi:10.1016/j.jelechem.2012.11.018}}.

\bibitem{Costa2013tqm}
R.~Costa, C.~M. Pereira, F.~Silva,
  \href{http://pubs.rsc.org/en/content/articlelanding/2013/ra/c3ra40584b}{Electric
  double layer studies at the interface of mercury{\textendash}binary ionic
  liquid mixtures with a common anion}, {RSC} Advances 3 (2013) 11697--11706.
\newblock \href {http://dx.doi.org/10.1039/C3RA40584B}
  {\path{doi:10.1039/C3RA40584B}}.
\newline\urlprefix\url{http://pubs.rsc.org/en/content/articlelanding/2013/ra/c3ra40584b}

\bibitem{Gomes2014txa}
C.~Gomes, R.~Costa, C.~M. Pereira, A.~F. Silva,
  \href{http://pubs.rsc.org/en/content/articlelanding/2014/ra/c4ra03977g}{The
  electrical double layer at the ionic {liquid/Au} and pt electrode interface},
  {RSC} Advances 4~(55) (2014) 28914--28921.
\newblock \href {http://dx.doi.org/10.1039/C4RA03977G}
  {\path{doi:10.1039/C4RA03977G}}.
\newline\urlprefix\url{http://pubs.rsc.org/en/content/articlelanding/2014/ra/c4ra03977g}

\bibitem{Costa2014vks}
R.~Costa, C.~M. Pereira, F.~Silva,
  \href{http://www.sciencedirect.com/science/article/pii/S0013468613022433}{Dicationic
  ionic liquid: Insight in the electrical double layer structure at mercury,
  glassy carbon and gold surfaces}, Electrochimica Acta 116 (2014) 306--313.
\newblock \href {http://dx.doi.org/10.1016/j.electacta.2013.11.034}
  {\path{doi:10.1016/j.electacta.2013.11.034}}.
\newline\urlprefix\url{http://www.sciencedirect.com/science/article/pii/S0013468613022433}

\bibitem{Loth2010}
M.~S. Loth, B.~Skinner, B.~I. Shklovskii, Anomalously large capacitance of an
  ionic liquid described by the restricted primitive model, Physical Review E
  82~(5) (2010) 056102.
\newblock \href {http://dx.doi.org/10.1103/PhysRevE.82.056102}
  {\path{doi:10.1103/PhysRevE.82.056102}}.

\bibitem{Vatamanu2010}
J.~Vatamanu, O.~Borodin, G.~D. Smith, Molecular insights into the potential and
  temperature dependences of the differential capacitance of a room-temperature
  ionic liquid at graphite electrodes, Journal of the American Chemical Society
  132~(42) (2010) 14825--14833.
\newblock \href {http://dx.doi.org/10.1021/ja104273r}
  {\path{doi:10.1021/ja104273r}}.

\bibitem{Dou2011}
Q.~Dou, M.~L. Sha, H.~Y. Fu, G.~Z. Wu, Molecular dynamics simulation of the
  interfacial structure of [c(n)mim][pf6] adsorbed on a graphite surface:
  effects of temperature and alkyl chain length, Journal of Physics-condensed
  Matter 23~(17) (2011) 175001.
\newblock \href {http://dx.doi.org/10.1088/0953-8984/23/17/175001}
  {\path{doi:10.1088/0953-8984/23/17/175001}}.

\bibitem{Kislenko2013wgo}
S.~A. Kislenko, R.~H. Amirov, I.~S. Samoylov,
  \href{http://iopscience.iop.org/1742-6596/418/1/012021}{Molecular dynamics
  simulation of the electrical double layer in ionic liquids}, Journal of
  Physics: Conference Series 418~(1) (2013) 012021--8.
\newblock \href {http://dx.doi.org/10.1088/1742-6596/418/1/012021}
  {\path{doi:10.1088/1742-6596/418/1/012021}}.
\newline\urlprefix\url{http://iopscience.iop.org/1742-6596/418/1/012021}

\bibitem{Vatamanu2014uro}
J.~Vatamanu, L.~Xing, W.~Li, D.~Bedrov, Influence of temperature on the
  capacitance of ionic liquid electrolytes on charged surfaces., Physical
  Chemistry Chemical Physics 16~(11) (2014) 5174--5182.
\newblock \href {http://dx.doi.org/10.1039/c3cp54705a}
  {\path{doi:10.1039/c3cp54705a}}.

\bibitem{Liu2014tvb}
X.~Liu, Y.~Han, T.~Yan, Temperature effects on the capacitance of an
  imidazolium-based ionic liquid on a graphite electrode: A molecular dynamics
  simulation, {ChemPhysChem} 15~(12) (2014) 2503--2509.
\newblock \href {http://dx.doi.org/10.1002/cphc.201402220}
  {\path{doi:10.1002/cphc.201402220}}.

\bibitem{Costa2015vkn}
R.~Costa, C.~M. Pereira, A.~Fernando~Silva,
  \href{http://linkinghub.elsevier.com/retrieve/pii/S1388248115001095}{Structural
  ordering transitions in ionic liquids mixtures}, Electrochemistry
  Communications 57 (2015) 10--13.
\newblock \href {http://dx.doi.org/10.1016/j.elecom.2015.04.012}
  {\path{doi:10.1016/j.elecom.2015.04.012}}.
\newline\urlprefix\url{http://linkinghub.elsevier.com/retrieve/pii/S1388248115001095}

\bibitem{Chen2017wlo}
M.~Chen, Z.~A.~H. Goodwin, G.~Feng, A.~A. Kornyshev,
  \href{http://www.sciencedirect.com/science/article/pii/S157266571730783X}{On
  the temperature dependence of the double layer capacitance of ionic liquids},
  Journal of Electroanalytical Chemistry\href
  {http://dx.doi.org/10.1016/j.jelechem.2017.11.005}
  {\path{doi:10.1016/j.jelechem.2017.11.005}}.
\newline\urlprefix\url{http://www.sciencedirect.com/science/article/pii/S157266571730783X}

\bibitem{Fedorov2014uyy}
M.~V. Fedorov, A.~A. Kornyshev, Ionic liquids at electrified interfaces,
  Chemical Reviews 114~(5) (2014) 2978--3036.
\newblock \href {http://dx.doi.org/http://dx.doi.org/10.1021/cr400374x}
  {\path{doi:http://dx.doi.org/10.1021/cr400374x}}.

\bibitem{Kornyshev2014vjv}
A.~A. Kornyshev, R.~Qiao,
  \href{http://dx.doi.org/10.1021/jp5047062}{Three-dimensional double layers},
  The Journal of Physical Chemistry C 118~(32) (2014) 18285--18290.
\newblock \href {http://dx.doi.org/10.1021/jp5047062}
  {\path{doi:10.1021/jp5047062}}.
\newline\urlprefix\url{http://dx.doi.org/10.1021/jp5047062}

\bibitem{Merlet2014vdh}
C.~Merlet, D.~T. Limmer, M.~Salanne, R.~van Roij, P.~A. Madden, D.~Chandler,
  B.~Rotenberg, \href{http://dx.doi.org/10.1021/jp503224w}{The electric double
  layer has a life of its own}, The Journal of Physical Chemistry C 118 (2014)
  18291--18298.
\newblock \href {http://dx.doi.org/10.1021/jp503224w}
  {\path{doi:10.1021/jp503224w}}.
\newline\urlprefix\url{http://dx.doi.org/10.1021/jp503224w}

\bibitem{Vu2012wkc}
T.~H. Vu, In situ {STM} studies at electrified au(hkl){\textbar}aqueous
  electrolyte and ionic liquid interfaces, Ph.D. thesis, University of Bern,
  Bern (2012).

\bibitem{Hess2008}
B.~Hess, C.~Kutzner, D.~van~der Spoel, E.~Lindahl, Gromacs 4: Algorithms for
  highly efficient, load-balanced, and scalable molecular simulation, Journal
  of Chemical Theory and Computation 4~(3) (2008) 435--447.
\newblock \href {http://dx.doi.org/10.1021/ct700301q}
  {\path{doi:10.1021/ct700301q}}.

\bibitem{Martinez2009}
L.~Mart\'{i}nez, R.~Andrade, E.~G. Birgin, J.~M. Mart\'{i}nez, Packmol: A
  package for building initial configurations for molecular dynamics
  simulations, Journal of Computational Chemistry 30~(13) (2009) 2157--2164.
\newblock \href {http://dx.doi.org/10.1002/jcc.21224}
  {\path{doi:10.1002/jcc.21224}}.

\bibitem{Essmann1995}
U.~Essmann, L.~Perera, M.~L. Berkowitz, T.~Darden, H.~Lee, L.~G. Pedersen, A
  smooth particle mesh ewald method, Journal of Chemical Physics 103~(19)
  (1995) 8577--8593.

\bibitem{Yeh1999}
I.~Yeh, M.~Berkowitz, {Ewald summation for systems with slab geometry}, Journal
  of Chemical Physics 111 (1999) 3155--3162.

\bibitem{Bussi2007}
G.~Bussi, D.~Donadio, M.~Parrinello, Canonical sampling through velocity
  rescaling, The Journal of Chemical Physics 126~(1) (2007) 014101.

\bibitem{Ivanistsev2015tvg}
V.~Ivani{\v s}t{\v s}ev, K.~Kirchner, K.~Karu, I.~Lage-Estebanez, M.~V.
  Fedorov, \href{https://github.com/vladislavivanistsev/NaRIBaS}{{NaRIBaS:} A
  scripting framework for computational modelling of Nanomaterials and Room
  Temperature Ionic Liquids in Bulk and Slab},
  {www.github.com/vladislavivanistsev/NaRIBaS}, 2017.
\newline\urlprefix\url{https://github.com/vladislavivanistsev/NaRIBaS}

\bibitem{Fedorov2008a}
M.~V. Fedorov, A.~A. Kornyshev, Ionic liquid near a charged wall: Structure and
  capacitance of electrical double layer, Journal of Physical Chemistry B
  112~(38) (2008) 11868--11872.

\bibitem{Holovko2001whu}
M.~Holovko, V.~Kapko, D.~Henderson, D.~Boda, On the influence of ionic
  association on the capacitance of an electrical double layer, Chemical
  Physics Letters 341~(3{\textendash}4) (2001) 363--368.
\newblock \href {http://dx.doi.org/10.1016/S0009-2614(01)00505-X}
  {\path{doi:10.1016/S0009-2614(01)00505-X}}.

\bibitem{Feng2011vvj}
G.~Feng, J.~Huang, B.~G. Sumpter, V.~Meunier, R.~Qiao, A “counter-charge
  layer in generalized solvents” framework for electrical double layers in
  neat and hybrid ionic liquid electrolytes, Physical Chemistry Chemical
  Physics 13 (2011) 14723--14734.
\newblock \href {http://dx.doi.org/10.1039/c1cp21428d}
  {\path{doi:10.1039/c1cp21428d}}.

\bibitem{Kirchner2013tss}
K.~Kirchner, T.~Kirchner, V.~Ivani\v{s}t\v{s}ev, M.~Fedorov,
  \href{http://www.sciencedirect.com/science/article/pii/S0013468613009547}{Electrical
  double layer in ionic liquids: Structural transitions from multilayer to
  monolayer structure at the interface}, Electrochimica Acta 110 (2013)
  762--771.
\newblock \href {http://dx.doi.org/10.1016/j.electacta.2013.05.049}
  {\path{doi:10.1016/j.electacta.2013.05.049}}.
\newline\urlprefix\url{http://www.sciencedirect.com/science/article/pii/S0013468613009547}

\bibitem{Ivanistsev2014wtw}
V.~Ivani\v{s}t\v{s}ev, M.~V. Fedorov, Interfaces between charged surfaces and
  ionic liquids: Insights from molecular simulations, The Electrochemical
  Society Interface 23~(1) (2014) 65--69.

\bibitem{Ivanistsev2015usp}
V.~Ivani\v{s}t\v{s}ev, K.~Kirchner, T.~Kirchner, M.~V. Fedorov,
  \href{http://iopscience.iop.org/0953-8984/27/10/102101}{Restructuring of the
  electrical double layer in ionic liquids upon charging}, Journal of Physics:
  Condensed Matter 27~(10) (2015) 102101.
\newblock \href {http://dx.doi.org/10.1088/0953-8984/27/10/102101}
  {\path{doi:10.1088/0953-8984/27/10/102101}}.
\newline\urlprefix\url{http://iopscience.iop.org/0953-8984/27/10/102101}

\bibitem{Ivanistsev2014tqg}
V.~Ivani\v{s}t\v{s}ev, S.~{O’Connor}, M.~V. Fedorov,
  \href{http://www.sciencedirect.com/science/article/pii/S1388248114002690}{Poly(a)morphic
  portrait of the electrical double layer in ionic liquids}, Electrochemistry
  Communications 48 (2014) 61--64.
\newblock \href {http://dx.doi.org/10.1016/j.elecom.2014.08.014}
  {\path{doi:10.1016/j.elecom.2014.08.014}}.
\newline\urlprefix\url{http://www.sciencedirect.com/science/article/pii/S1388248114002690}

\bibitem{Nishi2011vpv}
N.~Nishi, T.~Uruga, H.~Tanida, T.~Kakiuchi,
  \href{http://pubs.acs.org/doi/abs/10.1021/la200252z}{Temperature dependence
  of multilayering at the free surface of ionic liquids probed by x-ray
  reflectivity measurements}, Langmuir 27~(12) (2011) 7531--7536.
\newblock \href {http://dx.doi.org/10.1021/la200252z}
  {\path{doi:10.1021/la200252z}}.
\newline\urlprefix\url{http://pubs.acs.org/doi/abs/10.1021/la200252z}

\bibitem{Wallauer2013uvg}
J.~Wallauer, M.~Drüschler, B.~Huber, B.~Roling,
  \href{http://www.znaturforsch.com/ab/v68b/68b1143.htm}{The differential
  capacitance of ionic liquid {\textbar} metal electrode interfaces – a
  critical comparison of experimental results with theoretical predictions},
  Zeitschrift für Naturforschung B 68b (2013) 1143--1153.
\newblock \href {http://dx.doi.org/10.5560/ZNB.2013-3153}
  {\path{doi:10.5560/ZNB.2013-3153}}.
\newline\urlprefix\url{http://www.znaturforsch.com/ab/v68b/68b1143.htm}

\bibitem{Costa2015tae}
R.~Costa, C.~M. Pereira, A.~F. Silva,
  \href{http://www.sciencedirect.com/science/article/pii/S0013468615005010}{Charge
  storage on ionic liquid electric double layer: The role of the electrode
  material}, Electrochimica Acta 167 (2015) 421--428.
\newblock \href {http://dx.doi.org/10.1016/j.electacta.2015.02.180}
  {\path{doi:10.1016/j.electacta.2015.02.180}}.
\newline\urlprefix\url{http://www.sciencedirect.com/science/article/pii/S0013468615005010}

\bibitem{Oll2017wes}
O.~Oll, T.~Romann, C.~Siimenson, E.~Lust,
  \href{http://www.sciencedirect.com/science/article/pii/S1388248117301984}{Influence
  of chemical composition of electrode material on the differential capacitance
  characteristics of the ionic liquid{\textbar}electrode interface},
  Electrochemistry Communications 82~(Supplement C) (2017) 39--42.
\newblock \href {http://dx.doi.org/10.1016/j.elecom.2017.07.015}
  {\path{doi:10.1016/j.elecom.2017.07.015}}.
\newline\urlprefix\url{http://www.sciencedirect.com/science/article/pii/S1388248117301984}

\bibitem{Romann2014uqi}
T.~Romann, O.~Oll, P.~Pikma, H.~Tamme, E.~Lust,
  \href{http://www.sciencedirect.com/science/article/pii/S0013468614001455}{Surface
  chemistry of carbon electrodes in 1-ethyl-3-methylimidazolium
  tetrafluoroborate ionic liquid – an in situ infrared study}, Electrochimica
  Acta 125 (2014) 183--190.
\newblock \href {http://dx.doi.org/10.1016/j.electacta.2014.01.077}
  {\path{doi:10.1016/j.electacta.2014.01.077}}.
\newline\urlprefix\url{http://www.sciencedirect.com/science/article/pii/S0013468614001455}

\bibitem{Atkin2014tjb}
R.~Atkin, N.~Borisenko, M.~Drüschler, F.~Endres, R.~Hayes, B.~Huber,
  B.~Roling,
  \href{http://www.sciencedirect.com/science/article/pii/S0167732213002699}{Structure
  and dynamics of the interfacial layer between ionic liquids and electrode
  materials}, Journal of Molecular Liquids 192 (2014) 44--54.
\newblock \href {http://dx.doi.org/10.1016/j.molliq.2013.08.006}
  {\path{doi:10.1016/j.molliq.2013.08.006}}.
\newline\urlprefix\url{http://www.sciencedirect.com/science/article/pii/S0167732213002699}

\bibitem{Mezger2015tdg}
M.~Mezger, R.~Roth, H.~Schröder, P.~Reichert, D.~Pontoni, H.~Reichert,
  \href{http://scitation.aip.org/content/aip/journal/jcp/142/16/10.1063/1.4918742}{Solid-liquid
  interfaces of ionic liquid {solutions—Interfacial} layering and bulk
  correlations}, The Journal of Chemical Physics 142~(16) (2015) 164707.
\newblock \href {http://dx.doi.org/10.1063/1.4918742}
  {\path{doi:10.1063/1.4918742}}.
\newline\urlprefix\url{http://scitation.aip.org/content/aip/journal/jcp/142/16/10.1063/1.4918742}

\bibitem{Oll2016tjn}
O.~Oll, T.~Romann, P.~Pikma, E.~Lust,
  \href{http://www.sciencedirect.com/science/article/pii/S1572665716304131}{Spectroscopy
  study of ionic liquid restructuring at lead interface}, Journal of
  Electroanalytical Chemistry 778 (2016) 41--48.
\newblock \href {http://dx.doi.org/10.1016/j.jelechem.2016.08.016}
  {\path{doi:10.1016/j.jelechem.2016.08.016}}.
\newline\urlprefix\url{http://www.sciencedirect.com/science/article/pii/S1572665716304131}

\bibitem{Sebastian2015uzi}
P.~Sebastián, A.~P. Sandoval, V.~Climent, J.~M. Feliu,
  \href{http://www.sciencedirect.com/science/article/pii/S1388248115000806}{Study
  of the interface pt(111)/ [emmim][{NTf2]} using laser-induced temperature
  jump experiments}, Electrochemistry Communications 55 (2015) 39--42.
\newblock \href {http://dx.doi.org/10.1016/j.elecom.2015.03.012}
  {\path{doi:10.1016/j.elecom.2015.03.012}}.
\newline\urlprefix\url{http://www.sciencedirect.com/science/article/pii/S1388248115000806}

\bibitem{Kornyshev2007}
A.~A. Kornyshev, {Double-Layer in Ionic Liquids: Paradigm Change?}, Journal of
  Physical Chemistry B 111~(20) (2007) 5545--5557.
\newblock \href {http://dx.doi.org/10.1021/jp067857o}
  {\path{doi:10.1021/jp067857o}}.

\end{thebibliography}

\end{document}